 [2005/06/22 5.2/AAS markup document class]%
\RequirePackage{lineno}
\documentclass[apj]{emulateapj}
\usepackage{amssymb}
\usepackage{amsmath}
\usepackage{epsfig}
\usepackage{amsmath}
\usepackage{natbib}
\usepackage{commath}
\usepackage{color}
\usepackage{epstopdf}
\usepackage{hyperref}

\makeindex
\makeatletter
\let\o@verbatim\verbatim
\def\verbatim{%
  \ifhmode\unskip\par\fi
  \ifx\@currsize\normalsize
     \small
  \fi
  \o@verbatim
}

\renewcommand \verbatim@font {%
  \normalfont \ttfamily
  \catcode`\<=\active
  \catcode`\>=\active
}

\RequirePackage{shortvrb}
\MakeShortVerb{\|}

\begingroup
  \catcode`\<=\active
  \catcode`\>=\active
  \gdef<{\@ifnextchar<\@lt\@meta}
  \gdef>{\@ifnextchar>\@gt\@gtr@err}
  \gdef\@meta#1>{\m{#1}}
  \gdef\@lt<{\char`\<}
  \gdef\@gt>{\char`\>}
\endgroup
\def\@gtr@err{%
   \ClassError{ltxguide}{%
      Isolated \protect>%
   }{%
      In this document class, \protect<...\protect>
      is used to indicate a parameter.\MessageBreak
      I've just found a \protect> on its own.
      Perhaps you meant to type \protect>\protect>?
   }%
}
\def\verbatim@nolig@list{\do\`\do\,\do\'\do\-}

\newcommand{\m}[1]{\mbox{\it #1}}

\def\cmd#1{\cs{\expandafter\cmd@to@cs\string#1}}
\def\cmd@to@cs#1#2{\char\number`#2\relax}
\DeclareRobustCommand\cs[1]{\texttt{\char`\\#1}}
\begin{document}
\author{
IceCube Collaboration:
M.~G.~Aartsen\altaffilmark{1},
K.~Abraham\altaffilmark{2},
M.~Ackermann\altaffilmark{3},
J.~Adams\altaffilmark{4},
J.~A.~Aguilar\altaffilmark{5},
M.~Ahlers\altaffilmark{6},
M.~Ahrens\altaffilmark{7},
D.~Altmann\altaffilmark{8},
K.~Andeen\altaffilmark{9},
T.~Anderson\altaffilmark{10}
% ...
}
\altaffiltext{1}{Department of Physics, University of Adelaide, Adelaide, 5005, Australia}
\altaffiltext{2}{Physik-department, Technische Universit\"at M\"unchen, D-85748 Garching, Germany}
\altaffiltext{3}{DESY, D-15735 Zeuthen, Germany}
\altaffiltext{4}{Dept.~of Physics and Astronomy, University of Canterbury, Private Bag 4800, Christchurch, New Zealand}
\altaffiltext{5}{Universit\'e Libre de Bruxelles, Science Faculty CP230, B-1050 Brussels, Belgium}
\altaffiltext{6}{Dept.~of Physics and Wisconsin IceCube Particle Astrophysics Center, University of Wisconsin, Madison, WI 53706, USA}
\altaffiltext{7}{Oskar Klein Centre and Dept.~of Physics, Stockholm University, SE-10691 Stockholm, Sweden}
\altaffiltext{8}{Erlangen Centre for Astroparticle Physics, Friedrich-Alexander-Universit\"at Erlangen-N\"urnberg, D-91058 Erlangen, Germany}
\altaffiltext{9}{Department of Physics, Marquette University, Milwaukee, WI, 53201, USA}
\altaffiltext{10}{Dept.~of Physics, Pennsylvania State University, University Park, PA 16802, USA}
% ...

\title{Search for Sources of High Energy Neutrons with Four Years of Data from the IceTop Detector}

\begin{abstract}
	IceTop is an air shower array located on the Antarctic ice sheet at the geographic South Pole.
	IceTop can detect an astrophysical flux of neutrons from Galactic sources as an excess of cosmic ray air showers arriving from the source direction.
	Neutrons are undeflected by the Galactic magnetic field and can typically travel 10 ($E$ / PeV) pc before decay.
	Two searches are performed using 4 years of the IceTop dataset to look for a statistically significant excess of events with energies above 10 PeV ($10^{16}$ eV) arriving within a small solid angle.
	The all-sky search method covers from -90$^{\circ}$ to approximately -50$^{\circ}$ in declination. No significant excess is found.
	A targeted search is also performed, looking for significant correlation with candidate sources in different target sets.
	This search uses a higher energy cut (100 PeV) since most target objects lie beyond 1 kpc.
	The target sets include pulsars with confirmed TeV energy photon fluxes and high-mass X-ray binaries.
	No significant correlation is found for any target set.
	Flux upper limits are determined for both searches, which can constrain Galactic neutron sources and production scenarios.
\end{abstract}
\maketitle
%\tableofcontents

%\setpagewiselinenumbers
%\linenumbers
\section{Introduction}
\label{sec:intro}
The Galactic magnetic field (GMF) strongly affects the arrival distribution of charged cosmic rays, thereby obscuring their sources.
A compact source of high energy neutrons would manifest as a point source in cosmic ray arrival directions since neutrons are not deflected by magnetic fields.
Secondary neutral particles are an expected signature of hadronic acceleration in Galactic sources.
Neutral particles would be produced as the cosmic ray protons and nuclei undergo $pp$ and $p\gamma$ collisions, and photodisintegration, respectively, on the ambient photons and cosmic rays within the dense environment surrounding their source (see, e.g., \citep{2002APh....17...23C, Crocker:2005bb, 2006APh....26...41C, 2007PhRvD..75f3001A}).
For example, neutrons result from charge-exchange interactions,
\begin{equation*}
p\gamma \rightarrow n \pi^{+}
\end{equation*}
where a $\pi^{+}$ emerges with the proton's positive charge and the neutron retains most of the energy.
For interacting proton primaries, photons resulting from $\pi^{0}$ decays take a small fraction of the proton energy.
The production of neutrons exceeds the production of photons at the same energy \citep{Crocker:2005bb}.% if the proton primary energy spectrum falls as $E^{-\gamma}$ with a spectral index $\gamma>2$.
%An arrival direction search performed at or near the detector angular resolution may be sensitive enough to detect this signal above the local background.
%An energy-dependent effective range of $\approx$10 $E_{PeV}$ pc exists due to the neutron finite lifetime which motivates looking at higher energies since plausible accelerators are no closer than hundreds of pc to kpc.

It is plausible that known Galactic sources could produce high energy neutron fluxes, based on the measured TeV energy photon flux.
For some Galactic sources, the energy flux of TeV photons is greater than 1 eV cm$^{-2}$ s$^{-1}$ \citep{hinton2009}.
Sources producing particle fluxes with an $E^{-2}$ differential energy spectrum inject equal energy into each energy decade.
If sources in the Galaxy produce PeV photons in addition to TeV photons, the PeV photon energy flux would also exceed 1 eV cm$^{-2}$ s$^{-1}$ at Earth.
For sources that produce neutrons by hadronic processes as well, the neutron energy flux would be even higher since the neutron production rate exceeds the photon production rate, as noted previously.

Free neutrons undergo beta decay with a $880.0\pm0.9$ second half-life \citep{PDG:2014}.
Due to this decay, sources will only be visible within about 10 ($E$ / PeV) pc of Earth.
%This limits distances to potential bright sources to within $\approx$10 ($E$ / PeV) pc of Earth.
Since plausible accelerators such as young pulsars are no closer than 100 pc, searches at energies above 10 PeV are the most promising.

A diffuse flux of neutrons could be expected from interactions of cosmic ray primaries with ambient photons and the interstellar medium.
However, at PeV energies this flux would appear all over the sky since the effective range is less than the thickness of the Galactic disk.
This complicates a search for correlations with the Galactic plane since an excess signal could not be constrained to a particular region of the sky, for example Galactic latitudes $\abs{b}<10^{\circ}$.

At energies above $10^{18}$ eV (1 EeV), the Pierre Auger Observatory recently performed a search for neutrons in the Southern hemisphere finding no significant signal excesses or correlations with catalogs of Galactic objects, and established flux upper limits \citep{Aab:2012bha, Aab:2014caa}.
The Telescope Array experiment has established flux limits for point sources above 0.5 EeV in the Northern hemisphere \citep{2015ApJ...804..133A}.
%An analysis of SUGAR data reported an excess of events near the Galactic Center \citep{Clay:1992}, although this was not later confirmed with independent data.
KASCADE \citep{Antoni:2004sc} and CASA-MIA \citep{Chantell1997, Borione1998} found no point sources in the Northern hemisphere, also setting flux limits (an all-sky limit in the case of KASCADE).
AGASA \citep{Hayashida1999} and a re-analysis \citep{Bellido2001} of SUGAR data reported slight excesses towards the Galactic center, although these were later not confirmed by Auger \citep{Aab:2014yba}.

This paper reports the results of two searches for point-like signals in the arrival direction distribution of four years of IceTop data.
The two searches are an all-sky search for general hotspots on the sky and a search for correlations with nearby known Galactic sources.
In the all-sky search, we look for an excess of events from any direction in the sky, evaluating the significance of any excess using the method of Li and Ma \citep{Li:1983fv}.
The observable signature of a neutron flux is an excess of proton-like air showers.
The targeted search is treated as a stacked analysis using a set of candidate sources from an astrophysical catalog.
It is assumed that many or all of the candidates for a given set are emitting neutrons, so the combined signal should be more significant than that of a single target.
In both the all-sky and targeted searches, we set flux upper limits using the procedures of Feldman and Cousins \citep{Feldman:1997qc}.

This paper is organized as follows.
In Section \ref{sec:IC-IT}, the IceTop detector is described.
Section \ref{sec:recmethdata} summarizes the reconstruction methods and characteristics of the dataset.
The analysis methods and details of the search methods are described in Section \ref{sec:searches}.
The search results are presented in Section \ref{sec:results}.
A discussion of the results (Section \ref{sec:summary}) concludes the paper.

\section{IceCube / IceTop}
\label{sec:IC-IT}
IceTop is the surface air shower array of the IceCube Neutrino Observatory at the geographical South Pole located 2835 m above sea level \citep{IceCube:2013nn}.
Its final configuration consists of 81 stations covering 1 km$^{2}$ with an average station separation of 125 m.
Detector construction started in 2005 and finished in 2010.
A single station consists of two light-tight tanks separated by 10 m.
Each tank is 1.8 m in diameter, 1.3 m in height, and filled with transparent ice to a height of 0.9 m.
A tank contains two optical sensors, each consisting of a 10-inch Hamamatsu photomultiplier tube together with electronic boards for detection, digitization, and readout \citep{DOMpaper2009, PMTpaper2010}.
The two sensors are operated at different gains for increased dynamic range.
The IceTop trigger condition requires at least three stations to have recorded hits within a 5 $\mu$s time window \citep{IceCube:2013nn}.
IceTop detects showers at a rate of approximately 30 Hz with a minimum primary particle energy threshold of about 400 TeV.
Its surface location near the shower maximum makes it sensitive to the full electromagnetic component of the shower in addition to the muonic component.

%IceTop data are analyzed on-site and reduced according to physics-motivated event selections due to limited satellite bandwidth for data transmission to the Northern hemisphere.
Cosmic ray reconstruction relies on the optical detection of Cherenkov radiation within tanks of ice emitted by secondary particles produced by cosmic ray interactions in the upper atmosphere.
Information from individual tanks, including position, deposited charge, and pulse timing, is used to infer the air shower direction, core location, and shower size estimate $S_{125}$ which is related to the cosmic ray primary energy \citep{Aartsen:2013wda}.
%This data collection necessitates a prescale, which changes year-to-year with growing detector configurations.
%All showers that trigger eight or more stations are never prescaled and provide a consistent dataset.
%This analysis uses only these events.

Snow accumulates on the top of stations with time, attenuating the electromagnetic portion of the shower, lowering $S_{125}$.
This accumulation occurs in a non-uniform way due to wind patterns around nearby structures.
Snow depth measurements for each tank are performed twice a year allowing for depth interpolation at the time of an event.
An exponential correction factor is applied during event reconstruction to the signal of each tank such that the corrected tank signal $S_{125} = S_{125}^{\mbox{snow}}\ \mbox{exp}\left(x/ \lambda_{\mbox{eff}}\right)$.
Here, $S_{125}^{\mbox{snow}}$ is the detected signal in the tank, $x$ is the slant depth through the snow above the tank, and $\lambda_{\mbox{eff}}$ is the effective attentuation length due to the snow.
Values for $\lambda_{\mbox{eff}}$ are selected such that the resulting $S_{125}$ distributions for each year are consistent.
The attenuation length changes over time as the snow depth generally increases across the entire array \citep{IceTopICRC2015-628}.

\section{Reconstruction Methods and Dataset}
\label{sec:recmethdata}
This analysis uses four years of IceTop experimental data collected between May 2010 and May 2014.
For the first year of data (IC79), 73 stations were deployed; for each of the remaining 3 years (IC86), IceTop operated in its final 81-station configuration.

Event reconstructions are performed using the standard IceTop reconstruction method \citep{IceCube:2013nn}.
The values for the snow attenuation length $\lambda_{\mbox{eff}}$ differ for each year and are listed in Table \ref{tbl:dataset}. %: IC79 2.1 m, IC86-1 2.25 m, IC86-2 2.25 m, IC86-3 2.3 m.
The shower core location on the ground is determined by a signal-weighted likelihood fit to the shower front, with a typical resolution better than 10 m at the highest energies.
The primary arrival direction is determined from a fit to the arrival time distributions of signals in the tanks.
The angular resolution is the space angle that includes 68\% of reconstructed events that would arrive from a fixed direction.
%Due to the power-law nature of the cosmic ray energy spectrum, the dataset is dominated by events with energies just greater than a given energy threshold.
This value varies between 0.2$^{\circ}$ and 0.8$^{\circ}$ depending on energy and primary mass \citep{IceTopICRC2015-795}.
Above 10 PeV, the typical angular resolution, defined as the angle from the true event direction that contains 68\% of reconstructed event directions, is better than 0.5$^{\circ}$, which is taken as the representative value in the analysis.

The shower size estimate $S_{125}$ is determined by fitting the tank signals for the expected signal at 125 m from the shower core location.
The relationship between $S_{125}$ and primary cosmic ray energy is determined by comparison with Monte Carlo simulations for zenith angles less than $37^{\circ}$\citep{IceTopICRC2015-795}.
The energy resolution above 2 PeV is better than 0.1 in log$_{10}$ of the energy \citep{IceCube:2013nn}.

Events are selected by requiring a good fit to the shower lateral distribution, reconstructed core location lying within 400 m of the array center (not near the array boundary), and a cut on zenith angle within 37$^{\circ}$.
Requiring the reconstructed cores within 400 m yields a fiducial area $A=5.02\ 10^{5}\ \mbox{m}^{2}$.
For the final event selection for the all-sky search, we select energies above 10 PeV, and 100 PeV for the targeted search, resulting in 1,233,487 and 12,558 events, respectively.
The total livetime is 1363.8 days.
Table \ref{tbl:dataset} lists the livetime, number of events for each energy threshold, and effective snow attenuation length for each year.
%The higher energy cut for the targeted search allows for an increased range for which to include possible correlation candidates, as discussed in Section \ref{label:catalogs}.
\begin{table}[h]
\begin{center}
\caption{Detector configurations and their respective number of events and effective snow attenuation lengths for all years used in this analysis.\label{tbl:dataset}}
\begin{tabular}{ccccc}
\tableline
Configuration & Livetime & Number of Events & Snow Depth\\
& (days) & $N_{>10\ \mbox{PeV}}$ ($N_{>100\ \mbox{PeV}}$) & (meters) \\
\tableline
IC79   & 327.3 & 291,738  (2986) & 2.1 \\
IC86-1 & 342.0 & 305,138  (3173) & 2.25 \\
IC86-2 & 332.3 & 306,868  (3025) & 2.25 \\
IC86-3 & 362.2 & 329,743  (3374) & 2.3 \\
\tableline
Total & 1368.8 & 1,233,487  (12,558) & \nodata \\
\tableline
\end{tabular}
\end{center}
\end{table}

The targeted search uses a higher energy cut since most astrophysical objects of interest for this search lie at Galactic distances of order 1 kpc or greater.
This cut is also motivated by the fact the lower energy neutrons will not typically survive from 1 kpc and that lower energy contains only background contributions.
%This cut value allows to remove more isotropic background than signal, for example, due to different spectral indices of background and signal.

\section{Search Methods}
\label{sec:searches}
For both search methods, top-hat search windows are drawn on the sky.
This procedure allows for selecting events using a hard cut on the space angle between the event direction and the window center.
The locations of these search windows are described in the following sections with more detailed information about the two searches.
The radius of the search window in both searches is based on the actual IceTop point-spread function and is chosen such that it optimizes the sensitivity to a point source.
Point source sensitivity is optimized by choosing a window size $\chi$ based on the angular resolution.
The point spread function is taken to be $p\left(\theta\right) = (\theta/\sigma^{2})\ \mathrm{exp}\left(-\theta^{2}/2\sigma^{2}\right)$, where $\sigma = \psi/1.51$.
Here, $\theta$ is the space angle between the reconstructed and true arrival directions and $\psi$ is the angular resolution.
Using top-hat search windows, the sensitivity is optimized with $\chi = 1.59\sigma = 1.05\psi$, or $0.52^{\circ}$.

To find a signal excess within a search window, one must first know the expected number of events without signal, i.e., the background expectation value.
%Due to detector effects, for example nonuniform exposure to different parts of the sky and gaps in the detector uptime, the background expectation is not isotropic.
The background value for each search window is determined by time-scrambling the dataset many times.
Each time-scrambled set has the same number of events as the dataset.
For each event, we keep its zenith and azimuth angles in detector coordinates and randomly select another time in the dataset within a 24 hour window centered on the time of the event.
%This preserves the zenith and azimuth distributions of the data.
The search window content of the background expectation map is taken as the mean content of $10^{3}$ and $10^{6}$ time-scrambled maps for the all-sky and targeted searches, respectively.

\subsection{All-sky Search}
In the all-sky search, we look for excesses within search windows located in all parts of the sky within the field-of-view of IceTop.
These windows are centered on the pixels of a high-resolution HEALPix \citep{Gorski:2004by} map.
$N_{\mbox{side}}$ is a parameter used to define and generate the map's pixels, with higher values generating higher resolution maps.
We select a map defined by $N_{\mbox{side}}=128$ which provides 19,800 points within the IceTop field-of-view and simply provides central locations from which to draw the search windows.
The typical spacing between adjacent window locations in this map is $0.46^{\circ}$.
Although window overlap will cause correlations between neighboring windows, this ensures that all events are counted.
The data is first binned using a HEALPix map (``bin map'') with higher resolution ($N_{\mbox{side}}=256$) than the search window map.
The content of a given search window is the sum of contents of those pixels in the bin map whose centers fall within the search window.
The summed content of a search window is labelled $n$ ($n_{b}$) for the dataset (background).

Statistical significance of signals within search windows is based on the observed number of events $n$ and the background expectation value $n_{b}$.
The significance value of a given search window is calculated using the Li-Ma method \citep{Li:1983fv} shown in Eq. \ref{eq:lima},
\begin{equation}
\label{eq:lima}
\small{ S=\frac{n-n_{b}}{\abs{n-n_{b}}}\sqrt{2}\left(n\ \mathrm{ln}\left(\frac{n+\alpha n}{n_{b}+\alpha n}\right)+\frac{n_{b}}{\alpha} \mathrm{ln}\left(\frac{n_{b}+\alpha n_{b}}{n_{b}+\alpha n}\right)\right)^{1/2} }
\end{equation}
\noindent where we have replaced the Li-Ma parameters $N_{\mbox{on}}$ and $N_{\mbox{off}}$ with $n$ and $n_{b}/\alpha$, respectively.
The Li-Ma method is used only for the all-sky-search.
Typically $\alpha$ is the ratio of time spent observing on-source to the time spent observing an equivalent off-source solid angle.
Here, the parameter $\alpha$ is taken to be the ratio $n_{b}/\xi$, where $\xi$ is the sum of the contents of all search windows lying within $\pm90^{\circ}$ in right ascension and $\pm0.52^{\circ}$ in declination of the search window of interest, excluding the content value $n_{b}$ of the search window itself.
This definition of $\alpha$ provides a local estimate of $N_{\mbox{off}}$ for each search window.
IceTop observes large-scale anisotropy in cosmic ray arrival directions for energies above roughly 1 PeV \citep{IceTopICRC2015-373}; for example, a large deficit in the cosmic ray arrival direction distribution is observed from $30^{\circ}$ to $120^{\circ}$ in right ascension.
The estimate of $N_{\mbox{off}}$ should be representative of the expected cosmic ray flux in the vicinity of the search window, so this definition for $\alpha$ eliminates bias due to averaging over the field-of-view.

\subsection{Targeted Search}
\label{sec:targetedsearch}
The targeted search is performed to look for correlations of event directions with known nearby Galactic objects.
%Search windows with radii of $0.52^{\circ}$ are centered on the objects themselves.
We calculate the Poisson probability $p(n,n_{b})$ for observing $n$ or more events within the search window expecting $n_{b}$ for each object.
Fisher's method \citep{fisher1925} combines a set of independent probabilities to determine a single measure of significance $P_{F}$ for the set.
For a sequence of p-values $p_{1}, p_{2}, ... , p_{n}$, their product is $\pi=\prod_{i=1}^{n}p_{i}$.
Fisher's method allows to calculate the chance probability that a product $\pi$ of $n$ p-values obtained uniformly randomly would be less than or equal to the product $\pi_\text{obs}$ of the $n$ p-values observed:  $P_F(\pi \le \pi_\text{obs})$.
%Fisher's method allows to calculate the chance probability that repeated measurements of these p-values would yield a probability $P=\prod_{i=1}^{n}p_{i}$ less than or equal to $\pi$, or $P_{F}(P \leq \pi)$.

A supplemental measure of significance $P_{G}$ is provided by Good's method \citep{good1955} which allows for weights to be assigned to each probability.
In a similar way to Fisher's method, for a sequence of p-values $p_{i}$ with weight $w_{i}$, the weighted product $\pi_{w}=\prod_{i=1}^{n}p_{i}^{w_{i}}$.
Good's method allows to calculate the chance probability that a product $\pi_{w}$ of $n$ p-values obtained uniformly randomly with weights $w_{i}$ would be less than or equal to the product $\pi_\text{w,obs}$ of the $n$ p-values observed:  $P_G(\pi_{w} \le \pi_\text{w,obs})$.
%The chance probability $P_{G}(P_{w} \leq \pi_{w})$ is the weighted product of the p-values is not greater than $\pi_{w}$.
Here, these weights are proportional to the object's recorded electromagnetic flux listed in the catalog, its relative exposure to IceTop, and an expected flux attenuation factor.
This factor is equal to the survival probability for a neutron with energy equal to the median energy of an $E^{-2}$ energy spectrum between 100 PeV and 1 EeV to arrive from the distance of a candidate source object.
The weights are normalized such that their sum is 1 for each target set.

Treating the un-weighted and weighted probabilities ($P_{F}$, $P_{G}$) as individual test statistics, we calculate the fraction of time-scrambled datasets with corresponding values less than that observed with the data.
This post-trials fraction is an unbiased indicator of the correlation probability between the dataset and each source set.
Both the weighted and un-weighted probabilities and corresponding post-trials fractions are reported.
The un-weighted probability is independent of the assumption that neutron emission is proportional to the electromagnetic  emission and in how the flux, relative exposure, and decay probability are used to construct the object weight.

\subsection{Target Catalogs}
\label{label:catalogs}
We consider three distinct classes: millisecond pulsars \citep{2005AJ....129.1993M} (msec), $\gamma$-ray pulsars \citep{2013ApJS..208...17A} ($\gamma$-ray), and high mass X-ray binaries \citep{Liu:2007} (HMXB).
The msec catalog\footnote{http://www.atnf.csiro.au/research/pulsar/psrcat.} provides a comprehensive list of rotation-powered pulsars.
The $\gamma$-ray catalog is the second Fermi-LAT pulsar catalog.
The HMXB catalog\footnote{http://heasarc.gsfc.nasa.gov/w3browse/all/hmxbcat.html.} represents a comprehensive selection of X-ray sources, comprised of a compact object orbiting a massive OB class star.
These classes are considered candidate sources due to their independent evidence for high energy particle production and high flux measured at Earth.
The Galactic center lies outside the zenith angle cut and lies well beyond the effective neutron range even at energies of a few hundred PeV and is not considered a candidate in this search.

%% CHARACTERISTICS OF FERMI CATALOG
\begin{deluxetable*}{cccccccc}
\tablecolumns{8}
\tablewidth{0pc}
\tablecaption{Characteristics of the Fermi $\gamma$-ray catalog.\label{tbl:char-fermi}}
\tablehead{\colhead{Object Name} & \colhead{R.A.} & \colhead{Dec.} & \colhead{Distance} & \colhead{Energy Flux between} & \colhead{Relative} & \colhead{Survival} & \colhead{Normed} \\
      & \colhead{(\degr)} & \colhead{(\degr)} & \colhead{(kpc)} & \colhead{0.1-100 GeV (erg/cm$^2$/s)} &\colhead{Exposure} &\colhead{Probability\tablenotemark{a}} &\colhead{Weight} }
\startdata
J0101-6422	&15.30	&-64.38	&0.55	&1.047e-11	&0.902	&0.72	&0.026 \\
J1016-5857	&154.09	&-58.95	&2.9		&5.444e-11	&0.857	&0.18	&0.032 \\
J1028-5819	&157.12	&-58.32	&2.33	&2.426e-10	&0.851	&0.248	&0.199 \\
J1048-5832	&162.05	&-58.53	&2.74	&1.958e-10	&0.853	&0.194	&0.126 \\
J1105-6107	&166.36	&-61.13	&4.98	&4.89e-11		&0.876	&0.0509	&0.008 \\
J1112-6103	&168.06	&-61.06	&12.2	&2.034e-11	&0.875	&$<$\ 0.001	&$<$\ 0.001 \\
J1119-6127	&169.81	&-61.46	&8.4		&7.148e-11	&0.879	&0.0066	&0.002 \\
J1124-5916	&171.16	&-59.27	&4.8		&6.168e-11	&0.860	&0.057	&0.012 \\
J1125-5825	&171.43	&-58.42	&2.62	&8.9e-12		&0.852	&0.209	&0.006 \\
J1357-6429	&209.26	&-64.49	&2.5		&3.388e-11	&0.903	&0.22	&0.027 \\
J1410-6132	&212.59	&-61.53	&15.6	&2.63e-11		&0.879	&$<$\ 0.001	&$<$\ 0.001 \\
J1418-6058	&214.68	&-60.97	&1.6		&3.017e-10	&0.874	&0.38	&0.39 \\
J1420-6048	&215.03	&-60.80	&5.61	&1.698e-10	&0.873	&0.035	&0.020 \\
J1509-5850	&227.36	&-58.85	&2.62	&1.273e-10	&0.856	&0.209	&0.088 \\
J1513-5908	&228.48	&-59.14	&4.21	&3.243e-11	&0.858	&0.0807	&0.009 \\
J1531-5610	&232.87	&-56.18	&2.09	&1.94e-12		&0.831	&0.287	&0.002 \\
J1658-5324	&254.66	&-53.40	&0.93	&2.893e-11	&0.803	&0.57	&0.052 \\
\enddata
\tablenotetext{a}{Calculated using the median energy of an $E^{-2}$ spectrum between 100 PeV and 1 EeV}
\end{deluxetable*}

%% CHARACTERISTICS OF MSEC CATALOG
\begin{deluxetable*}{cccccccc}
\tablecaption{Characteristics of the msec catalog.\label{tbl:char-msec}}
\tablehead{\colhead{Object Name} & \colhead{R.A.} & \colhead{Dec.} & \colhead{Distance} & \colhead{Energy Flux at Sun} & \colhead{Relative} & \colhead{Survival} & \colhead{Normed} \\
      & \colhead{(\degr)} & \colhead{(\degr)} & \colhead{(kpc)} & \colhead{(erg/kpc$^2$/s)} &\colhead{Exposure} &\colhead{Probability\tablenotemark{a}} &\colhead{Weight} }
\startdata
J1017-7156	&154.46	&-71.94	&0.26	&1e+35	&0.951	&0.86	&0.5 \\
B0021-72F	&6.02	&-72.08	&4		&8.8e+33	&0.951	&0.09	&0.005 \\
J1125-6014	&171.48	&-60.24	&1.94	&2.3e+33	&0.868	&0.314	&0.004 \\
J1910-5959A	&287.93	&-59.97	&4.5		&1.6e+32	&0.866	&0.068	&$<$\ 0.001 \\
J1103-5403	&165.89	&-54.06	&3.16	&3.7e+32	&0.809	&0.151	&$<$\ 0.001 \\
J1216-6410	&184.03	&-64.17	&1.71	&4.9e+32	&0.900	&0.360	&0.001 \\
J1933-6211	&293.39	&-62.20	&0.63	&8.3e+33	&0.885	&0.69	&0.032 \\
J1740-5340A	&265.19	&-53.68	&3.4		&1.2e+34	&0.806	&0.13	&0.008 \\
J2129-5721	&322.34	&-57.35	&0.4		&9.9e+34	&0.842	&0.8		&0.4 \\
J1431-5740	&217.76	&-57.67	&4.07	&2.2e+32	&0.845	&0.0877	&$<$\ 0.001 \\
J0711-6830	&107.98	&-68.51	&1.04	&3.3e+33	&0.931	&0.537	&0.010 \\
J1629-6902	&247.29	&-69.05	&1.36	&9.9e+32	&0.934	&0.443	&0.003 \\
J2236-5527	&339.22	&-55.46	&2.03	&2.8e+32	&0.824	&0.297	&$<$\ 0.001 \\
J1757-5322	&269.31	&-53.37	&1.36	&8e+32	&0.803	&0.443	&0.002 \\
J1435-6100	&218.83	&-61.02	&3.25	&1.1e+32	&0.875	&0.143	&$<$\ 0.001 \\
J1337-6423	&204.38	&-64.38	&6.3		&2.9e+31	&0.902	&0.023	&$<$\ 0.001 \\
\enddata
\tablenotetext{a}{Calculated using the median energy of an $E^{-2}$ spectrum between 100 PeV and 1 EeV}
\end{deluxetable*}

%% CHARACTERISTICS OF HMXB CATALOG
\begin{deluxetable*}{cccccccc}
\tablecaption{Characteristics of the HMXB catalog.\label{tbl:char-hmxb}}
\tablehead{\colhead{Object Name} & \colhead{R.A.} & \colhead{Dec.} & \colhead{Distance} & \colhead{Energy Flux between} & \colhead{Relative} & \colhead{Survival} & \colhead{Normed} \\
      & \colhead{(\degr)} & \colhead{(\degr)} & \colhead{(kpc)} & \colhead{2-10 keV ($\mu$Jy)} &\colhead{Exposure} &\colhead{Probability\tablenotemark{a}} &\colhead{Weight} }
\startdata
1H0739-529			&116.85	&-53.33	&0.52	&0.7		&0.802	&0.73	&0.007 \\
1H0749-600			&117.57	&-61.10	&0.4		&0.7		&0.875	&0.8		&0.008 \\
GROJ1008-57			&152.44	&-58.29	&5		&1200	&0.851	&0.05	&0.9  \\
RXJ1037.5-5647		&159.40	&-56.80	&5		&3.3		&0.837	&0.05	&0.002 \\
1A1118-615			&170.24	&-61.92	&5		&0.1		&0.882	&0.05	&$<$\ 0.001 \\
4U1119-603			&170.31	&-60.62	&9		&10		&0.871	&0.005	&$<$\ 0.001 \\
IGRJ11215-5952		&170.44	&-59.86	&8		&42		&0.865	&0.008	&0.005 \\
2S1145-619			&177.00	&-62.21	&2.3		&4		&0.885	&0.25	&0.02 \\
1E1145.1-6141			&176.87	&-61.95	&8		&4		&0.883	&0.008	&$<$\ 0.001 \\
4U1223-624			&186.66	&-62.77	&3		&9		&0.889	&0.2		&0.02 \\
1H1249-637			&190.71	&-63.06	&0.3		&2.2		&0.891	&0.8		&0.03 \\
1H1253-761			&189.81	&-75.37	&0.24	&0.6		&0.968	&0.87	&0.008 \\
1H1255-567			&193.65	&-57.17	&0.11	&0.8		&0.840	&0.94	&0.01 \\
4U1258-61			&195.32	&-61.60	&2.4		&0.3		&0.880	&0.24	&0.001 \\
2RXPJ130159.6-635806	&195.50	&-63.97	&5.5		&6.3		&0.890	&0.037	&0.003 \\
SAXJ1324.4-6200		&201.11	&-62.01	&3.4		&0.4		&0.883	&0.1		&$<$\ 0.001 \\
2S1417-624			&215.30	&-62.70	&6		&2		&0.889	&0.03	&$<$\ 0.001 \\
SAXJ1452.8-5949		&223.21	&-59.82	&9		&0.045	&0.864	&0.005	&$<$\ 0.001 \\
XTEJ1543-568			&236.00	&-56.77	&10		&8		&0.836	&0.003	&$<$\ 0.001 \\
1H1555-552			&238.59	&-55.33	&0.96	&1.7		&0.822	&0.56	&0.013 \\
\enddata
\tablenotetext{a}{Calculated using the median energy of an $E^{-2}$ spectrum between 100 PeV and 1 EeV}
\end{deluxetable*}

\begin{figure}[h]
\begin{center}
\epsscale{1.6}
\plotone{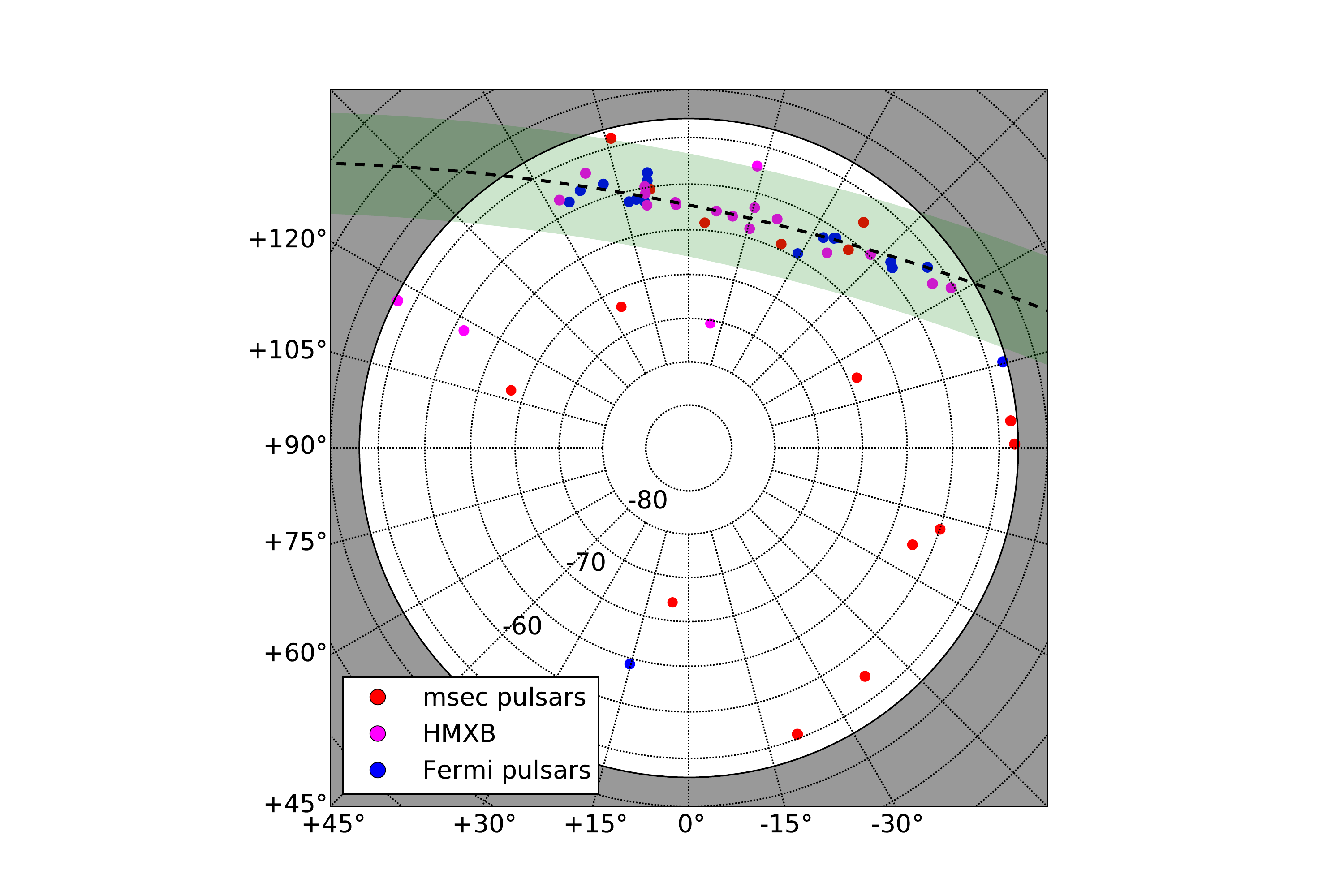}
\caption{Equatorial polar skymap of each catalog set.
The dashed black line indicates the Galactic plane and the green band shows $b = \pm 5^{\circ}$.
Each circle is 0.5$^{\circ}$ in radius.\label{fig:sourcemap}}
\end{center}
\end{figure}

Only objects with known distances are included in the final catalog selection.
Distances for each candidate are cross-checked with the TeVCat catalog\footnote{http://tevcat.uchicago.edu}.
Most objects are eliminated from each catalog by the zenith angle cut and by requiring that the distance is known.
Sources that further appear in multiple sets are retained only in the smaller set, resulting in 17 objects in the $\gamma$-ray set, 16 objects in the msec set, and 20 objects in the HMXB set as shown in Tables \ref{tbl:char-fermi}-\ref{tbl:char-hmxb} respectively.
The columns in each table are the object designation, right ascension, declination, distance, electromagnetic flux as recorded in the catalog, relative exposure value to IceTop, survival probability for a neutron with energy equal to the median energy of an $E^{-2}$ energy spectrum between 100 PeV and 1 EeV, and normalized weight value.
Figure \ref{fig:sourcemap} shows the locations of each object in equatorial coordinates.
The Galactic plane is depicted by a green band to illustrate the preferential association of the $\gamma$-ray pulsar and HMXB sets with that part of the sky.
\epsscale{1.0} % reset figure sizes to normal

%The targeted search uses a higher energy cut since most astrophysical objects of interest for this analysis lie at Galactic distances of order 1 kpc or greater.
%This cut also reduces the number of events so that any signal above the cut value is not masked by the much larger number of events with energies below 100 PeV.

%The zenith angle cut removes the Galactic center as a candidate target.
%The Galactic center is not considered as a candidate due to its large distance compared to the decay length even for 100 PeV neutrons.
%At 100 PeV, the expected survival distance is 1 kpc, whereas the half-thickness of the Galactic disk is on the order of 100 pc.
%The Galactic plane appear smeared across a large fraction of the sky

\subsection{Flux Upper Limit Calculation}
\label{sec:fulcalc}
Flux upper limits are calculated for both the all-sky and targeted searches using,
\begin{equation}
F_{UL} = 1.39\ s_{UL} / \zeta
\end{equation}
where $s_{UL}$ is the upper limit on the number of signal events in the search window and $\zeta=TA\mbox{cos}(\theta)\epsilon$ is the exposure of IceTop, where $T$ is the livetime, $A\ \mbox{cos}(\theta)$ is the projected detector area exposed to the search window which depends on the zenith angle $\theta$, and $\epsilon$ is the reconstruction efficiency (taken as 95\% according to Monte Carlo studies).
The signal upper limit $s_{UL}$ is calculated using a 90\% Feldman-Cousins confidence level \citep{Feldman:1997qc} based on $n$ and $n_{b}$ for the search window.
The factor 1.39 is a compensation factor to include signal events that fall outside the search window.
The search window includes only 71.8\% of signal events based on the top-hat window and the assumed IceTop point-spread function, therefore $s_{UL}$ is scaled by $1 / 0.718 = 1.39$.

The flux upper limit can be rewritten as, %in [$\mbox{km}^{-2}\ \mbox{yr}^{-1}$] is,
\begin{equation}
F_{UL} = 0.776\ \left( s_{UL}/\mbox{cos}(\theta)\right)\ [\mbox{km}^{-2}\ \mbox{yr}^{-1}],
\label{eq:particleflux}
\end{equation}
by substituting $TA\epsilon=1.79\ \mbox{km}^{2}\ \mbox{yr}$.
For an assumed $E^{-2}$ energy spectrum over the 100 PeV - 1 EeV energy decade, the median energy is 181.8 PeV.
The median energy flux upper limit in [$\mbox{eV}\ \mbox{cm}^{-2}\ \mbox{s}^{-1}$] over this energy range can be written as,
\begin{equation}
F_{UL}^{E} = 0.447\ \left( s_{UL}/\mbox{cos}(\theta)\right).
\label{eq:energyflux}
\end{equation}
Over the 10 PeV - 1 EeV energy decades, the median energy is 19.80 PeV so the conversion factor between the particle flux and median energy flux upper limits is,
\begin{equation}
1\ \mbox{part.}\ \mbox{km}^{-2}\ \mbox{yr}^{-1} = 0.0628\ \mbox{eV}\ \mbox{cm}^{-2}\ \mbox{s}^{-1}
\label{eq:conversionfactor}
\end{equation}

\begin{figure}[h]
\begin{center}
\epsscale{1.4}
\plotone{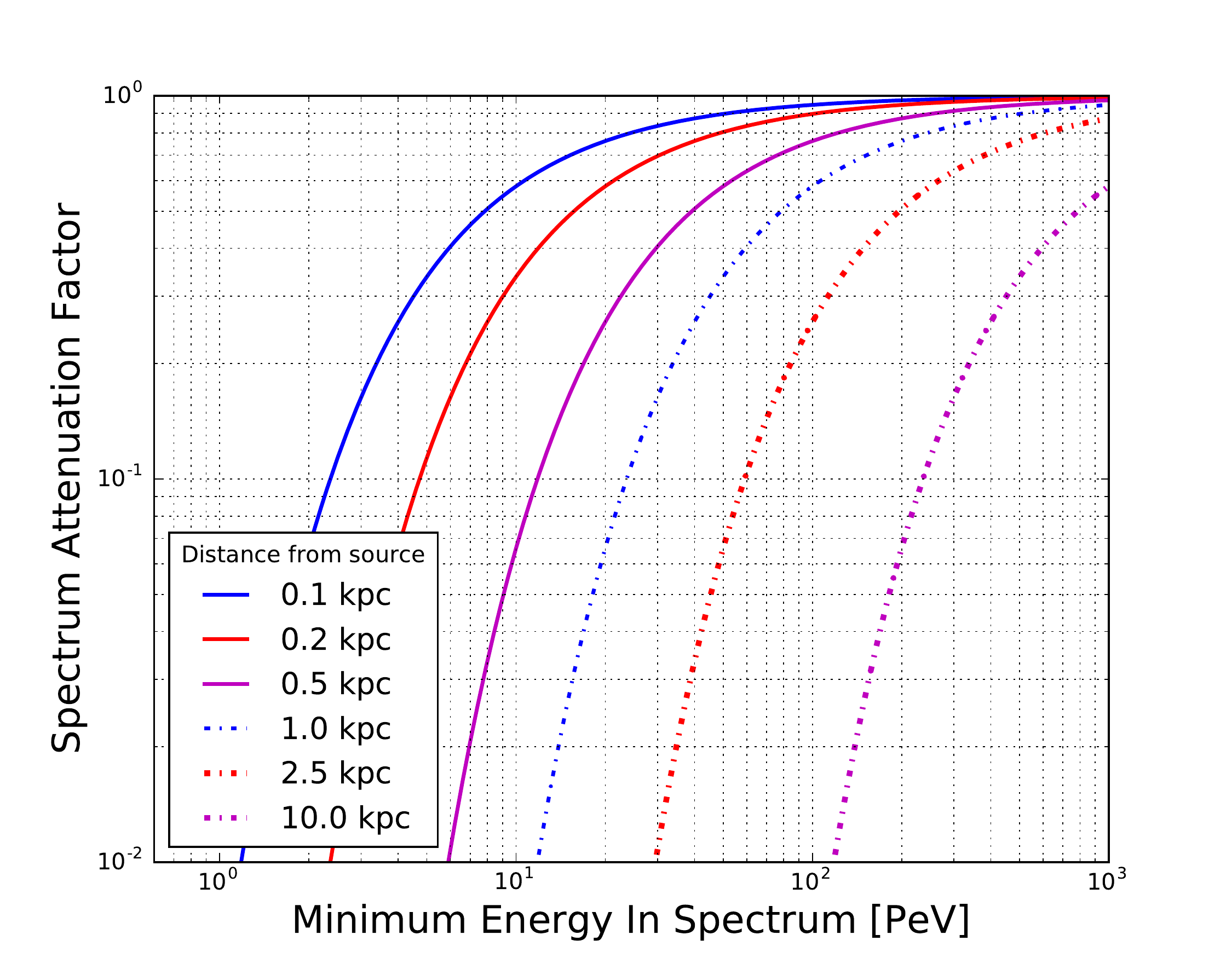}
\caption{Spectrum attentuation factor due to neutron decay as a function of minimum energy for an $E^{-2}$ spectrum and distance from the source.
The attenuation factor is a function of the median energy which itself depends on the minimum energy.\label{fig:decayatten}}
\end{center}
\end{figure}

An important point to note is that Eq. \ref{eq:energyflux} assumes an $E^{-2}$ energy spectrum \textit{as measured at Earth}, which is related to the source energy spectrum only after accounting for neutron decay factors that depend on the source distance.
Figure \ref{fig:decayatten} shows the attenuation factor of the energy spectrum injected at the source due to decay during propagation for representative distances.
For a sufficiently distant source, the source spectrum would be harder than that observed at Earth.
The lower energy portions of the spectrum are increasingly suppressed with distance as these neutrons are removed.
\begin{figure}[h]
\begin{center}
\epsscale{1.4}
\plotone{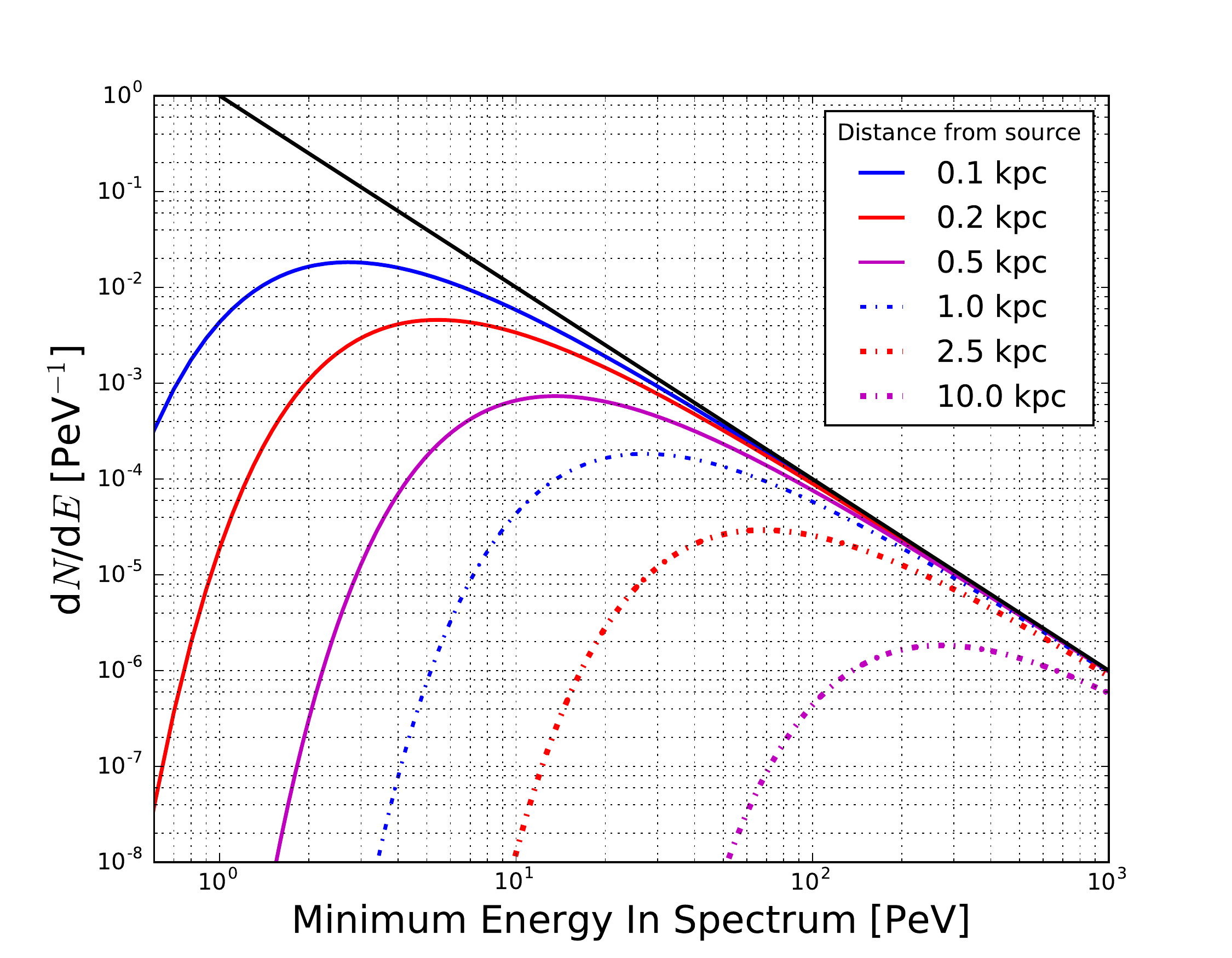}
\caption{Effect of decay attenuation on an $E^{-2}$ energy spectrum at source as a function of minimum energy and distance from the source.\label{fig:spectrumatten}}
\end{center}
\end{figure}
%The median energy conversion factors apply only to the observed spectrum.
\epsscale{1.0}

The attenuation curves in Figure \ref{fig:decayatten} have a strong effect on the sensitivity of the searches.
The all-sky search uses a 10 PeV energy threshold, thus it is sensitive only to sources at extremely close distances due to the large number of events lying near threshold.
The targeted search is sensitive mostly to higher neutron energies closer to EeV energies, which are capable of crossing larger Galactic distances.
For example, a suppression factor $S$ can be defined as the ratio between the number of neutrons with an injected $E^{-2}$ spectrum observed after including attenuation to the number observed not including attenuation for the same $E^{-2}$ spectrum.
For an $E^{-2}$ spectrum between 10 PeV and 1 EeV, removal of half the neutrons from the observed spectrum, or $S=0.5$, corresponds to a propagation distance of about 0.15 kpc.
Between 100 PeV and 1 EeV, $S=0.5$ corresponds to a distance of about 1.25 kpc.
Generally speaking, the sensitivity of any neutron search will be shifted towards the higher energy portion of the injected energy spectrum at the source due to decay, unless sources are sufficiently close that decay does not significantly modify the energy spectrum.
This can be seen in Figure \ref{fig:spectrumatten} which shows an example $E^{-2}$ energy spectrum modified by the distance-dependent decay attenuation.

These flux limits are time-averaged values based on the IceTop exposure $\zeta$.
Particularly for the objects in the targeted source sets, it is possible that transient fluxes may temporarily exceed these limits.
The energy flux limits derived from Eq. \ref{eq:energyflux} are strongly dependent on the assumption that an injected $E^{-2}$ energy spectrum at the source is not strongly modified in the energy range the limit applies to by neutron decay en-route.

\section{Results}
\label{sec:results}
\subsection{All-sky Search}
\begin{figure}[htb]
\begin{center}
\epsscale{1.4}
\plotone{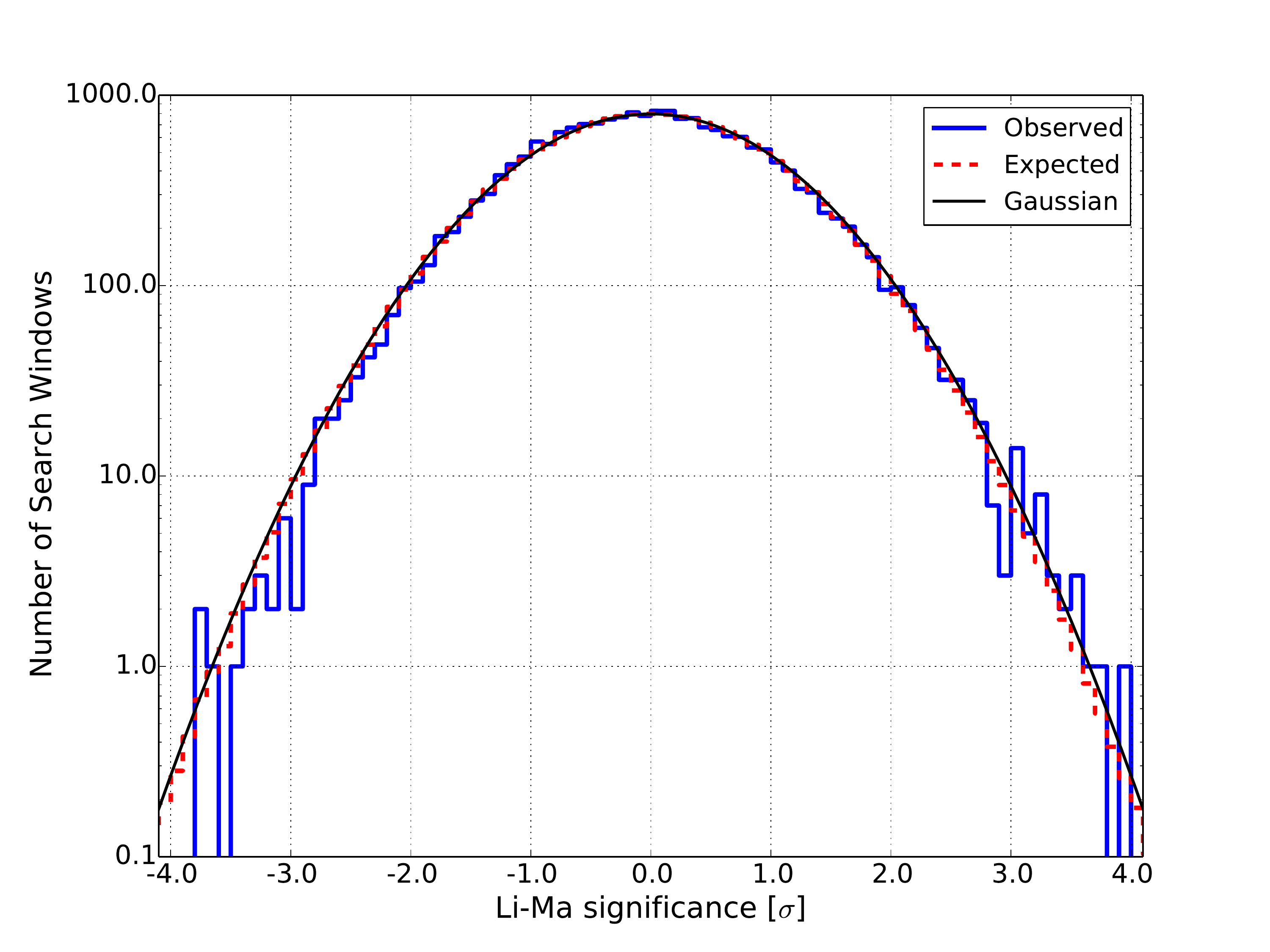}
\caption{Differential histograms of Li-Ma values (blue) and the isotropic expectation (green).\label{fig:histlima}
The dashed line shows the Gaussian approximation for the expected Li-Ma distribution in the case that deviations result only from statistical fluctuations.}
\end{center}
\end{figure}

\begin{figure}[htb]
\begin{center}
\epsscale{1.4}
\plotone{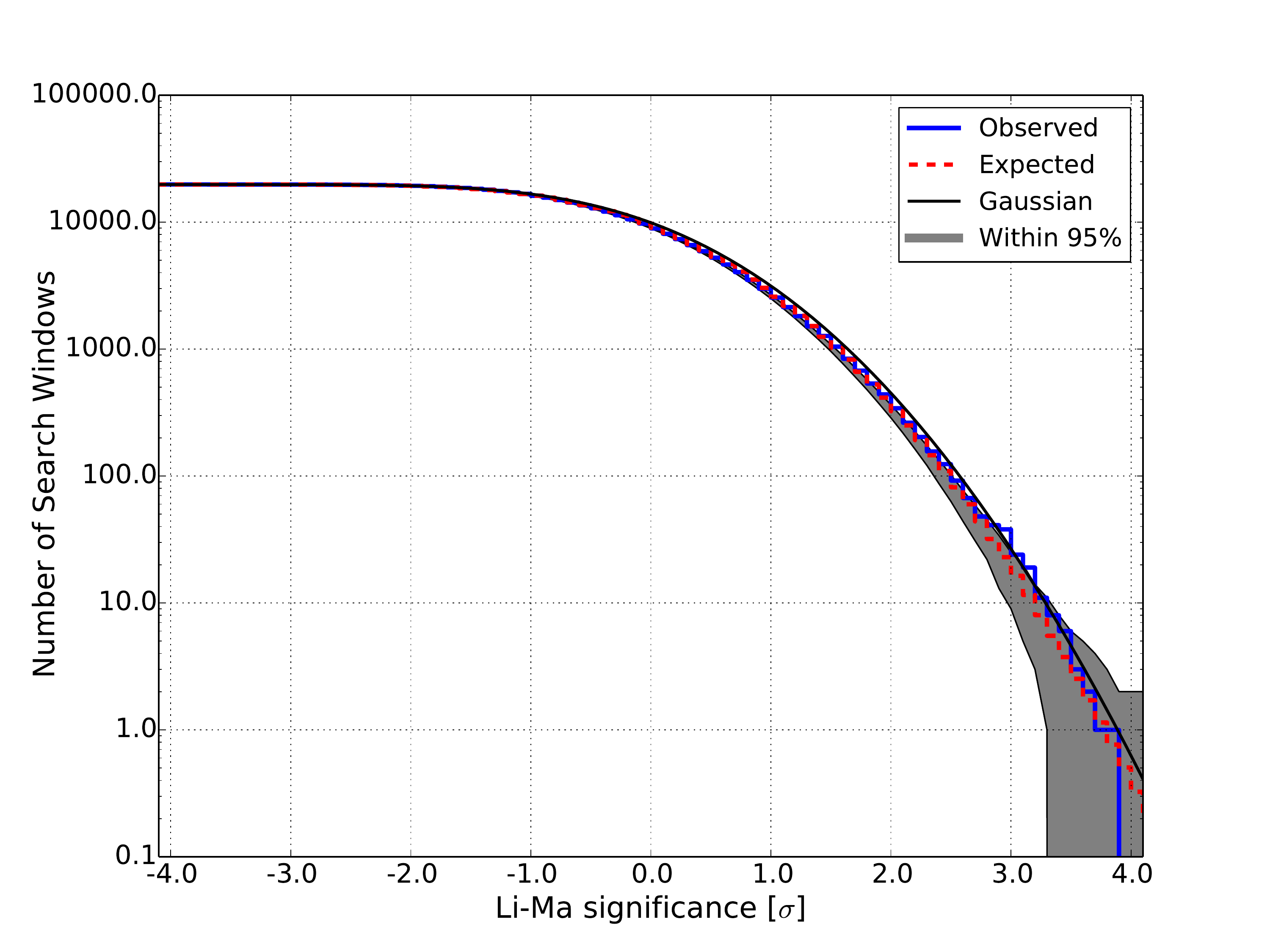}
\caption{Cumulative histograms of Li-Ma values (blue) and the isotropic expectation (green).\label{fig:histlimacdf}}
\end{center}
\end{figure}

Figures \ref{fig:histlima} and \ref{fig:histlimacdf} show the differential and cumulative distributions of the 19,800 Li-Ma values compared to the isotropic expectation.
In both figures, the blue and green lines show the Li-Ma significance distribution for the data and isotropy, respectively.
There are no Li-Ma values larger than 4.
The dashed line shows the Gaussian form expected for the distribution to follow if deviations from isotropy are due only to statistical fluctuations.
In Figure \ref{fig:histlimacdf}, the gray shaded region in the cumulative plot shows the 95\% containment band for isotropy; the presence of search windows with statistically significant signal excess would extend above and to the right of this band.
The absence of such a feature indicates that no statistically significant signal excess is observed and that the observed excesses are consistent with fluctuations about the expectation.

\begin{figure}[htb]
\begin{center}
\epsscale{1.2}
\plotone{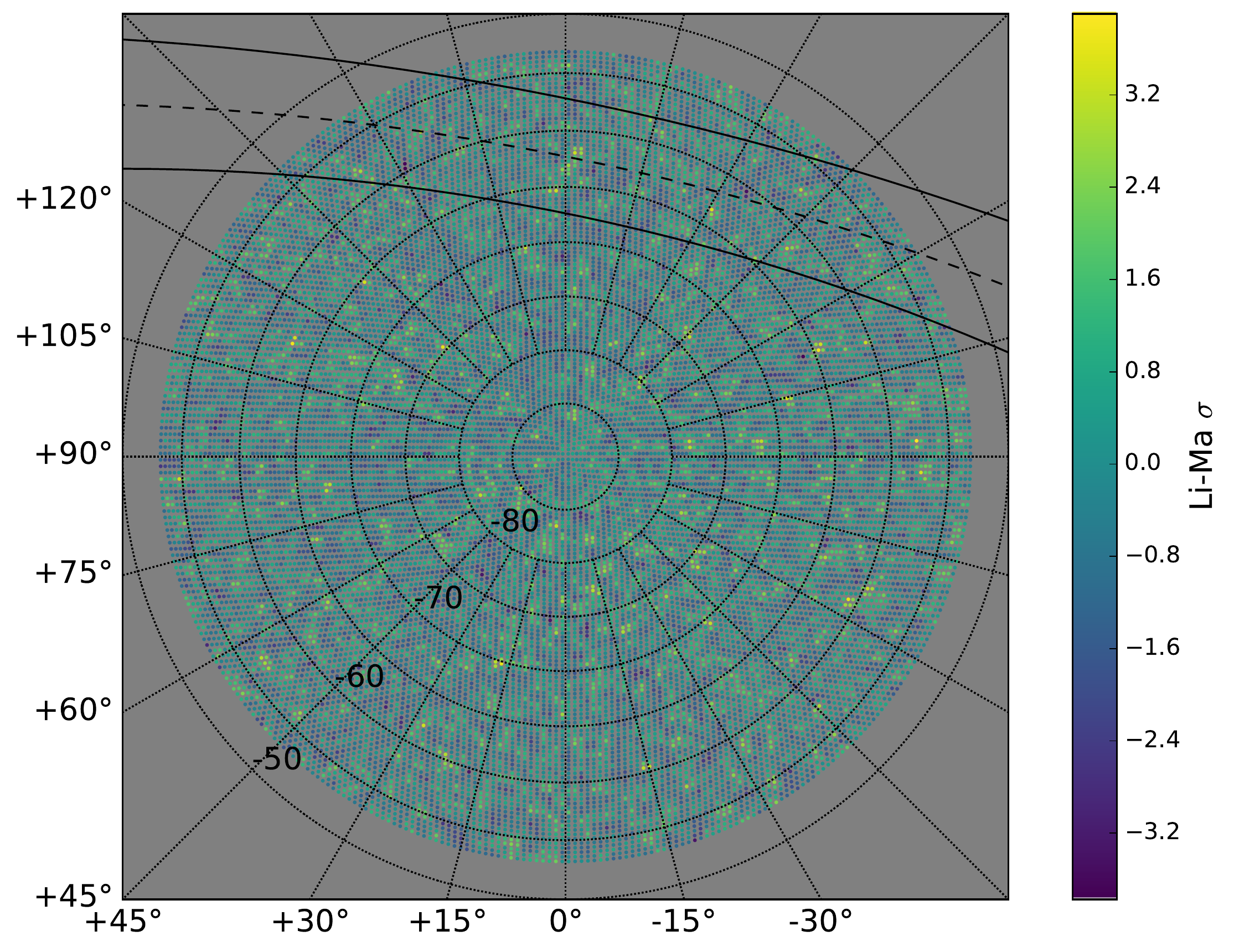}
\caption{Equatorial polar skymap of Li-Ma values for each search window.
The dashed black line indicates the Galactic plane and the solid black lines depict $b=\pm5^{\circ}$.\label{fig:limamap}}
\end{center}
\end{figure}

\begin{figure}[htb]
\begin{center}
\epsscale{1.2}
\plotone{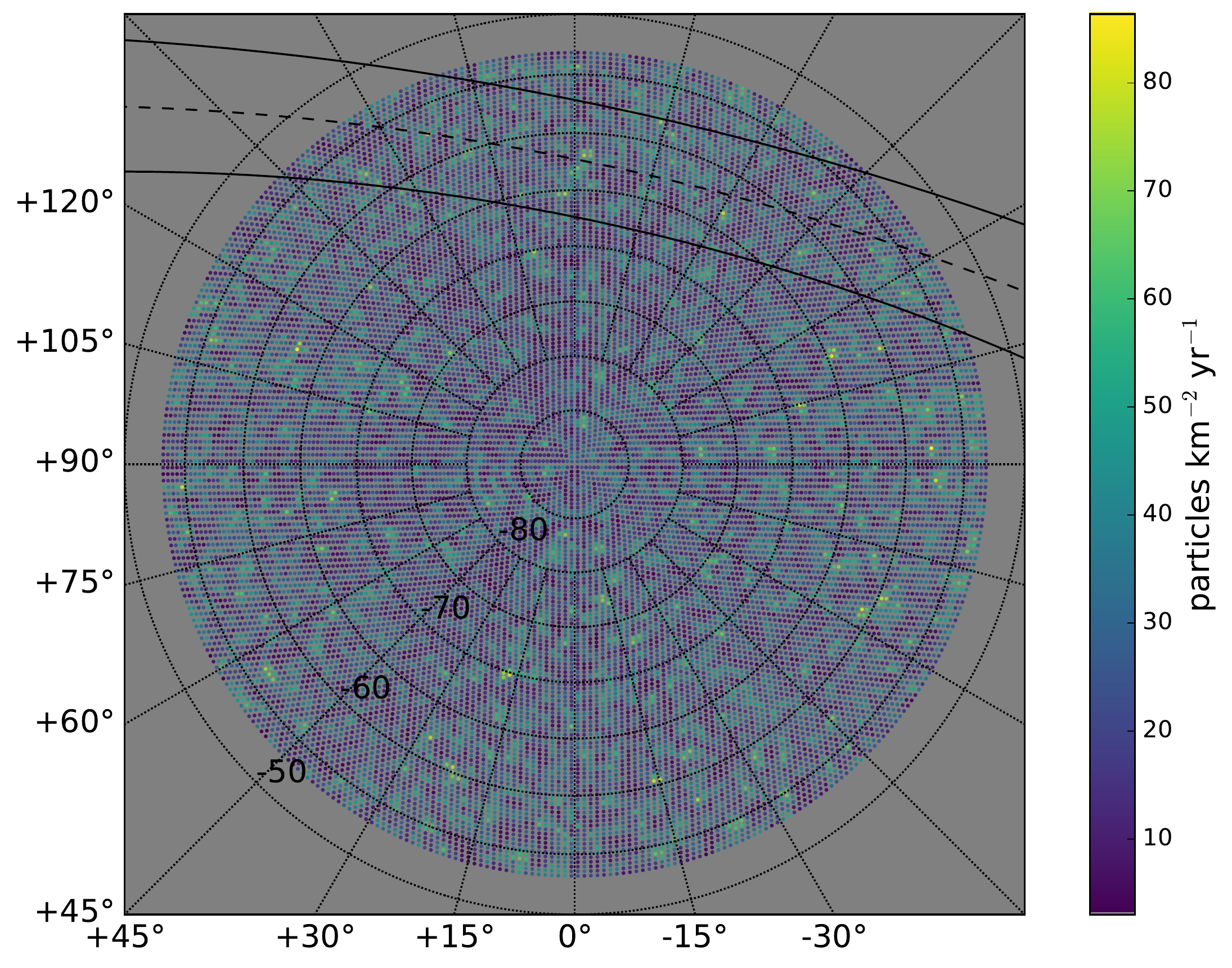}
\caption{Equatorial polar skymap of flux upper limit values for each search window.
The dashed black line indicates the Galactic plane and the solid black lines depict $b=\pm5^{\circ}$.
\label{fig:fluxmap}}
\end{center}
\end{figure}

Figures \ref{fig:limamap} and \ref{fig:fluxmap} show skymaps of the Li-Ma and flux upper limit values for each search window.
No statistically significant clustering on the sky is observed, including the Galactic plane depicted by the black dashed ($b=0^{\circ}$) and solid ($b=\pm5^{\circ}$) lines.
As noted previously, the energies of most events used in this search lie close to the 10 PeV energy cut, which corresponds to a neutron range of order 100 pc.
The sphere from which signal could arrive is contained within the Galactic disk so any excesses arising from cosmic ray interactions in the disk would be distributed over the entire field-of-view, not concentrated within a narrow band across the sky.

\begin{figure}[htb]
\begin{center}
\plotone{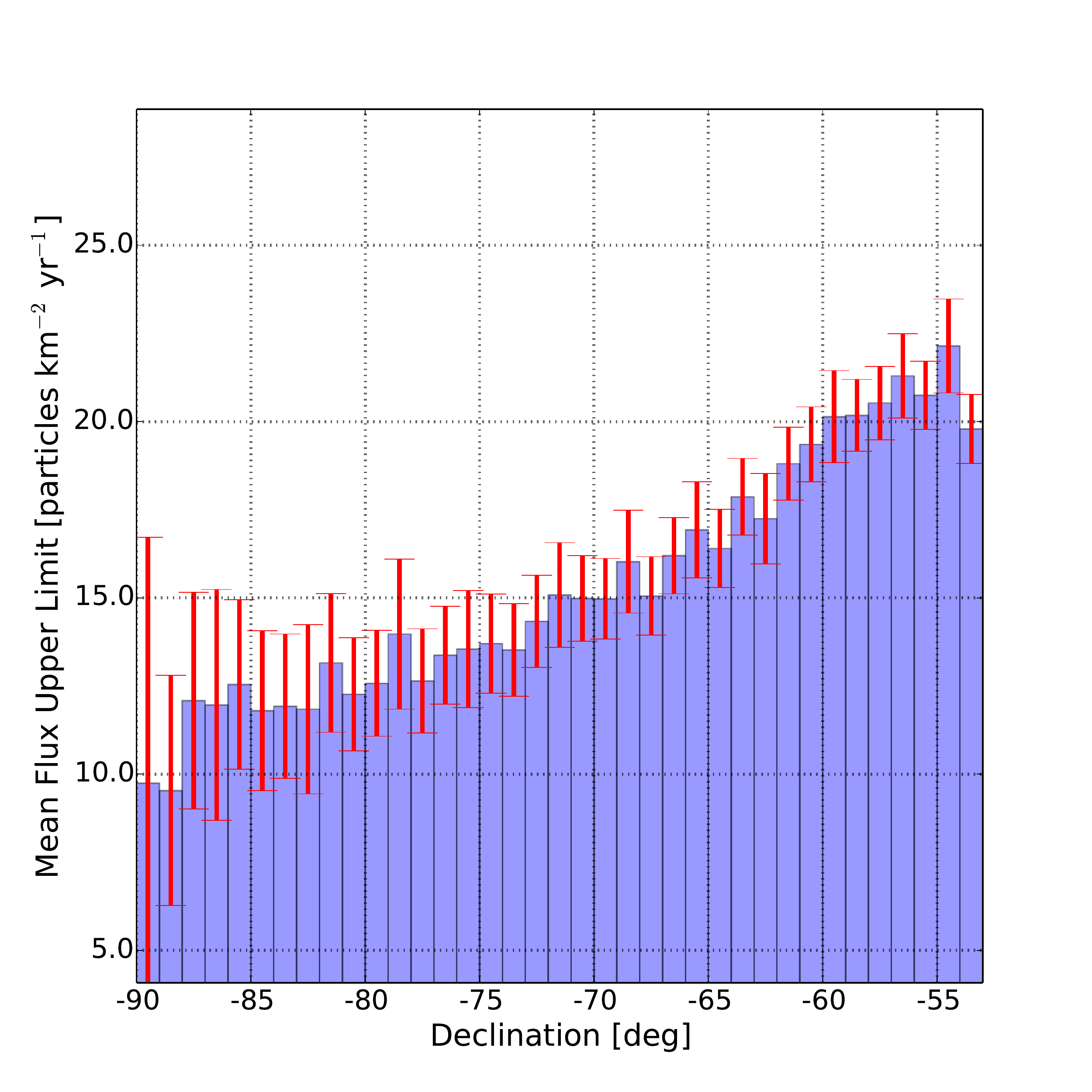}
\caption{Mean flux upper limit (90\% C.L.) for $1^{\circ}$ declination bins for the all-sky search.
The error bars indicate the statistical uncertainty on the mean value since there are many search windows within each declination band.\label{fig:fluxdecmap}}
\end{center}
\end{figure}

Figure \ref{fig:fluxdecmap} shows the mean flux upper limit as a function of declination for the all-sky search.
The limits are strongest near the South Pole due to the maximal exposure, but there is greater uncertainty on the mean since there are less search windows in declination bands closest to the pole.
\epsscale{1.4}
\epsscale{1.0}

\subsection{Targeted Search}
%% GENERAL CATALOG RESULTS
\begin{table}
\begin{center}
\caption{Targeted search results with each catalog.
Values in parentheses give the post-trials probability.\label{tbl:results-catalogs}}
\begin{tabular}{ccc}
\tableline
Catalog & Un-weighted $P_{F}$ & Weighted $P_{G}$ \\
\tableline
$\gamma$-ray & 0.999 (0.976) & 0.910 (0.776) \\
msec & 0.809 (0.408) & 0.888 (0.778) \\
HMXB & 0.999 (0.988) & 0.946 (0.971) \\
\tableline
\end{tabular}
\end{center}
\end{table}

%% MOST SIGNIFICANT OBJECT RESULTS

%% FERMI CATALOG RESULTS
\begin{deluxetable*}{cccccc}
\tablecaption{Targeted search results for the Fermi $\gamma$-ray catalog.\label{tbl:results-fermi}}
\tablehead{\colhead{Object Name} & \colhead{Observed} & \colhead{Background} & \colhead{$F_{UL}$} & \colhead{$F_{UL}^{E}$} & \colhead{Poisson} \\
   &\colhead{$n$}  &\colhead{Estimate $n_{b}$}  & \colhead{(km$^{-2}$ yr$^{-1}$)} & \colhead{(eV cm$^{-2}$ s$^{-1}$)} &\colhead{Probability $p$} }
\startdata
J1016-5857	&3	&2.62	&4.35	&2.51	&0.487 \\
J1028-5819	&1	&1.80	&2.44	&1.41	&0.835 \\
\textbf{J1048-5832}	&\textbf{5}	&\textbf{2.77}	&\textbf{6.57}	&\textbf{3.79}	&\textbf{0.147} \\
J1105-6107	&2	&3.79	&2.19	&1.26	&0.892 \\
J1112-6103	&3	&3.79	&3.29	&1.90	&0.729 \\
J1125-5825	&2	&2.65	&3.02	&1.74	&0.742 \\
J1124-5916	&2	&1.73	&3.78	&2.18	&0.517 \\
J1119-6127	&3	&2.71	&4.16	&2.40	&0.508 \\
J0101-6422	&3	&2.81	&3.98	&2.29	&0.534 \\
J1357-6429	&2	&2.34	&3.09	&1.78	&0.679 \\
J1410-6132	&1	&2.75	&1.79	&1.03	&0.936 \\
J1418-6058	&2	&2.81	&2.83	&1.63	&0.770 \\
J1420-6048	&2	&2.62	&2.98	&1.72	&0.737 \\
J1509-5850	&0	&2.37	&0.83	&0.48	&1.000 \\
J1513-5908	&0	&1.81	&1.07	&0.62	&1.000 \\
J1531-5610	&0	&2.78	&0.68	&0.39	&1.000 \\
J1658-5324	&1	&2.59	&2.06	&1.19	&0.925 \\
\enddata
\end{deluxetable*}

%% MSEC CATALOG RESULTS
\begin{deluxetable*}{cccccc}
\tablecaption{Targeted search results for the msec pulsar catalog.\label{tbl:results-msec}}
\tablehead{\colhead{Object Name} & \colhead{Observed} & \colhead{Background} & \colhead{$F_{UL}$} & \colhead{$F_{UL}^{E}$} & \colhead{Poisson} \\
   &\colhead{$n$}  &\colhead{Estimate $n_{b}$}  & \colhead{(km$^{-2}$ yr$^{-1}$)} & \colhead{(eV cm$^{-2}$ s$^{-1}$)} &\colhead{Probability $p$} }
\startdata
J0711-6830	&2	&2.54	&2.85	&1.64	&0.720 \\
J1103-5403	&2	&1.99	&3.77	&2.17	&0.591 \\
J1017-7156	&2	&2.23	&3.01	&1.74	&0.654 \\
J1125-6014	&2	&1.80	&3.68	&2.12	&0.537 \\
J1216-6410	&3	&2.67	&4.10	&2.36	&0.499 \\
B0021-72F	&4	&1.95	&5.42	&3.12	&0.133 \\
J1337-6423	&5	&3.02	&6.00	&3.46	&0.188 \\
J1435-6100	&1	&2.49	&1.94	&1.12	&0.917 \\
J1431-5740	&3	&1.84	&5.13	&2.96	&0.281 \\
J1629-6902	&0	&2.89	&0.57	&0.33	&1.000 \\
J2236-5527	&4	&2.72	&5.54	&3.19	&0.289 \\
\textbf{J1933-6211}	&\textbf{6}	&\textbf{3.20}	&\textbf{7.26}	&\textbf{4.18}	&\textbf{0.106} \\
J1910-5959A	&2	&3.21	&2.58	&1.49	&0.830 \\
J2129-5721	&1	&2.48	&2.04	&1.18	&0.916 \\
J1740-5340A	&1	&2.52	&2.10	&1.21	&0.919 \\
J1757-5322	&3	&2.18	&5.08	&2.93	&0.372 \\
\enddata
\end{deluxetable*}

%% HMXB CATALOG RESULTS
\begin{deluxetable*}{cccccc}
\tablecaption{Targeted search results for the HMXB catalog.\label{tbl:results-hmxb}}
\tablehead{\colhead{Object Name} & \colhead{Observed} & \colhead{Background} & \colhead{$F_{UL}$} & \colhead{$F_{UL}^{E}$} & \colhead{Poisson} \\
   &\colhead{$n$}  &\colhead{Estimate $n_{b}$}  & \colhead{(km$^{-2}$ yr$^{-1}$)} & \colhead{(eV cm$^{-2}$ s$^{-1}$)} &\colhead{Probability $p$} }
\startdata
1H0739-529			&0	&2.51	&0.82	&0.47	&1.000 \\
1H0749-600			&1	&2.44	&1.98	&1.14	&0.913 \\
GROJ1008-57			&1	&2.82	&1.82	&1.05	&0.941 \\
RXJ1037.5-5647		&2	&3.22	&2.66	&1.53	&0.832 \\
IGRJ11215-5952		&2	&1.95	&3.56	&2.05	&0.579 \\
4U1119-603			&2	&1.83	&3.63	&2.09	&0.546 \\
1A1118-615			&1	&2.08	&2.18	&1.26	&0.876 \\
1E1145.1-6141			&3	&2.64	&4.21	&2.43	&0.492 \\
2S1145-619			&3	&2.31	&4.49	&2.59	&0.408 \\
4U1223-624			&1	&3.39	&1.45	&0.84	&0.966 \\
1H1249-637			&1	&2.09	&2.14	&1.23	&0.877 \\
1H1253-761			&2	&3.41	&2.19	&1.26	&0.855 \\
1H1255-567			&2	&2.01	&3.61	&2.08	&0.598 \\
2RXPJ130159.6-635806	&2	&2.87	&2.71	&1.56	&0.781 \\
4U1258-61			&1	&2.93	&2.28	&1.31	&0.947 \\
SAXJ1324.4-6200		&1	&2.49	&1.93	&1.11	&0.917 \\
2S1417-624			&4	&2.86	&5.01	&2.89	&0.322 \\
\textbf{SAXJ1452.8-5949}		&\textbf{3}	&\textbf{1.69}	&\textbf{5.15}	&\textbf{2.97}	&\textbf{0.239} \\
XTEJ1543-568			&3	&2.81	&4.29	&2.47	&0.532 \\
1H1555-552			&0	&2.42	&8.35	&4.81	&1.000 \\
\enddata
\end{deluxetable*}

%% RIGHT ASCENSION ROTATION RESULTS
\begin{table}[htb]
\scriptsize
\begin{center}
\caption{Results of right ascension rotation tests for $\gamma$-ray pulsar and HMXB catalogs.
Values in parentheses give the post-trials probability.\label{tbl:results-ratest}}
\begin{tabular}{cccc}
\tableline
Probability & $+45^{\circ}$ & $+90^{\circ}$ & $+180^{\circ}$ \\
\tableline
HMXB $P_{F}$ & 0.070 (0.002) & 0.991 (0.828) & 0.978 (0.741) \\
HMXB $P_{G}$ & 0.262 (0.129) & 0.164 (0.116) & 0.946 (0.982) \\
$\gamma$-ray $P_{F}$ & 0.803 (0.408) & 0.126 (0.016) & 0.991 (0.866) \\
$\gamma$-ray $P_{G}$ & 0.074 (0.025) & 0.586 (0.311) & 0.845 (0.645) \\
\tableline
\end{tabular}
\end{center}
\end{table}
Table \ref{tbl:results-catalogs} lists the correlation probabilities for each catalog with the corresponding post-trials probability in parentheses.
No significant correlation is observed with any catalog.
Tables \ref{tbl:results-fermi}-\ref{tbl:results-hmxb} give details of each object. %, including the neutron flux $F_{UL}$ and median energy flux $F_{UL}^{E}$ upper limits above 100 PeV using Eqs. \ref{eq:particleflux} and \ref{eq:energyflux}.
The columns in each table are the object designation, observed number of events within the search window, background estimate in the window, particle flux above 100 PeV according to Eq. \ref{eq:particleflux}, energy flux above 100 PeV according to Eq. \ref{eq:energyflux}, and Poisson probability $p(n, n_{b})$ for observing $n$ events with an expectation number $n_{b}$.
These flux limits assume an $E^{-2}$ energy spectrum as measured at Earth.
The most significant object in each catalog is highlighted in bold.
%The post-trials probability for the minimum $p$ in each catalog, shown in parentheses in the table, also indicates that no evidence for excess PeV neutron flux from the candidates is observed.

\begin{figure}[htb]
\begin{center}
\epsscale{1.3}
\plotone{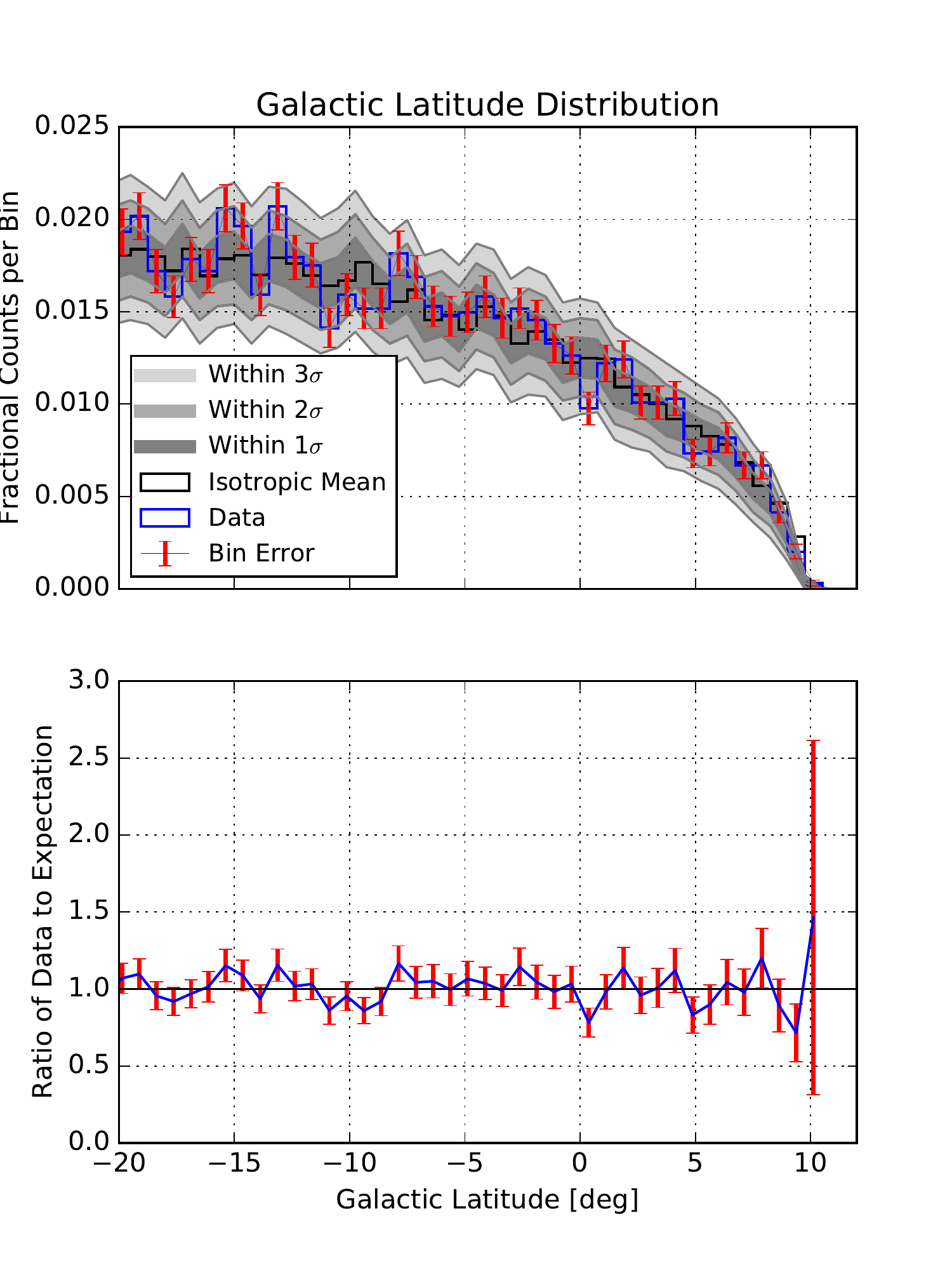}
\caption{Distribution of number of events above 100 PeV as a function of Galactic latitude.
In the top frame, the blue histogram shows the data; the black line shows the isotropic expectation from 10000 time-scrambled datasets.
The red error bars show the Poisson uncertainty in the data histogram.
The gray shaded bands depict the 68\%, 95\%, and 99\% containment bands for isotropy in each latitude bin.
The bottom frame shows the ratio between the data and isotropic expectation.\label{fig:galbdistmap}}
\end{center}
\end{figure}
\epsscale{1.0}
The catalog probabilities do not appear to be distributed uniformly between 0 and 1, since at least one probability value would be expected to lie below 0.809 in over 99.5\% of sets of 6 uniformly distributed random samples.
There exists an underfluctuation in the data along $b=0^{\circ}$ compared to the background expectation, as illustrated in Figure \ref{fig:galbdistmap}.
The preferential clustering of the $\gamma$-ray pulsar and HMXB catalogs along the Galactic plane combined with this underfluctuation acts to drive the catalog probabilities to higher values.
Since typically $n<n_{b}$, the individual Poisson $p$-values are close to 1.
This is checked by rotating these catalogs by a prescribed amount in right ascension and expecting lower catalog probabilities due to higher $n$ and similar $n_{b}$ in the windows.
These values are shown in Table \ref{tbl:results-ratest} for different rotation values.

We also note that there are four pairs of objects which lie within $1^{\circ}$ of each other.
In all four cases, the objects in each pair are distinct from each other, lie at different distances, and are from different catalogs.
We find consistent results with Table \ref{tbl:results-catalogs} when we mask the object with the farther distance.

\section{Summary and Discussion}
\label{sec:summary}
IceTop does not observe a statistically significant point source of cosmic ray arrival directions.
Using Equation \ref{eq:conversionfactor} the all-sky mean flux upper limits for individual declination bands correspond to energy fluxes between about 0.6 - 1.2 eV cm$^{-2}$ s$^{-1}$ between 100 PeV and 1 EeV assuming an $E^{-2}$ neutron energy spectrum as measured at Earth, which are comparable to TeV photon fluxes for Galactic objects \citep{hinton2009}.
These flux limits are the first neutron flux upper limits  in the Southern hemisphere for energies in the 10 PeV to 1 EeV energy decades.
Again, it is important to note that neutron decay en-route will modify the energy spectrum as illustrated in Figure \ref{fig:decayatten}, so the source spectrum would be generally softer than that constrained.
The limits in both searches are strongly dependent on the assumption that an injected $E^{-2}$ spectrum is not significantly modified by decay, as noted in Section \ref{sec:fulcalc}.
For the all-sky search, this restricts the applicability of the limits within a small volume around Earth.
For the targeted search, there are a number of objects that lie within 1 kpc, so their limits are most compatible with the base assumption.

%For parent protons with an $E^{-2}$ spectrum, these upper limits do not necessarily provide stringent constraints on the production origin of the TeV photons and the energy evolution of proton spectra.
As noted previously, hadronic production of photons by protons with an $E^{-2}$ spectrum will inject equal power into each energy decade, and the neutron production at least equals the photon production.
At present, these flux upper limits do not strongly constrain the TeV photon production mechanism, or the shape of the parent energy spectrum.
No significant correlation is found with known nearby Galactic objects characterized by GeV-TeV energy photon emission and plausibly capable of producing PeV neutrons.

%As an example of a possible expected neutron signal, we note that the most significant object found in the $\gamma$-ray catalog (J1048-5832) has an integrated photon flux of $2.457\ 10^{-7}$ photons cm$^{-2}$ s$^{-1}$ between 0.1 and 100 GeV with a best fit spectral index $\Gamma=1.6$ \citep{2013ApJS..208...17A}.
%Assuming this GeV photon spectrum extends up to and beyond 100 PeV without a cutoff and simplistic 1:1 production of high energy photons and neutrons, we would expect to observe about 2 neutrons with energies above 100 PeV from this object based on the IceTop exposure and a neutron survival probability of about 11\% from this object's distance.

The non-observation of PeV neutrons may simply indicate that these objects are not producing neutrons at these energies, or that typical Galactic neutron sources are not near Earth.
Local PeV neutron production in the Galaxy could simply be episodic or transient, for example, occurring during supernova explosions or other extremely high energy particle production events.
Alternatively, the sources may emit particle jets continuously, but their number may be few and the jets are not oriented towards the Earth.
Individual sources could emit weakly but be densely distributed.

Additionally, the environment around any sources may not be sufficiently dense to facilitate neutron production by cosmic ray interaction such that the primaries escape the acceleration region into interstellar space before interacting and producing neutrons.
In this case, neutrons decay in interstellar space relatively near the primary source producing secondary protons \citep{1997PhRvL..79.2616B}.
These secondary protons then propagate diffusively in the GMF, so sources that are sufficiently far away will not manifest a point source signal of cosmic ray neutrons, but could contribute to a proton signal that is smeared on the sky and not necessarily pointing back to the original source; this argument was presented by \citep{Bossa2003} when they considered EeV neutrons from the Galactic center.
At PeV energies, neutrons would penetrate much less into the surrounding medium, so any potential signal from the resulting protons would be strongly suppressed by the scattering effects of the GMF and masked by the background cosmic ray flux.
%On the other hand, neutron production may be high at lower energies where protons comprise a larger fraction, but the resulting lower energies severely restrict observations to within a few pc of the production site.

At higher energies, for example, between 10-100 PeV, this process could further enrich the cosmic ray proton fraction above that which is directly accelerated at the source.
The knee in the cosmic ray spectrum is observed around 4 PeV which is interpreted as an indication of a maximum attainable rigidity of typical Galactic cosmic ray sources and of associated changes in elemental composition (see e.g., \citep{Horandel:2005bb, Blasi:2014roa}).
It is plausible that the maximum attainable energy for the proton energy spectrum in nearby sources may not extend well above the knee energy although for heavier compositions this scales with the nuclear charge $Z$.
Above 10 PeV, the cosmic ray flux becomes progressively heavier with energy and with a decreasing proton fraction which is roughly 20\% at 10 PeV \citep{IceTopICRC2015-795, Apel:2013dga}.
This suggests that such secondary enrichment may be unlikely since a recovering proton fraction is not observed at energies between roughly 10 PeV to a few 100 PeV.
%At ultra-high energies, interpretation of depth of air shower maxima suggests that between about $10^{17.8}-10^{18.3}$ eV, the cosmic ray flux is composed of predominantly light nuclei \citep{Aab:2014kda}.

The non-observation of a PeV neutron flux does not necessarily preclude the existence of a PeV photon flux.
The neutron energy spectrum at lower energies becomes increasingly modified by decay.
PeV photons, on the other hand, have an absorption length considerably larger than the neutron decay distance and will maintain an unmodified energy spectrum that more resembles the injected spectrum at source.
PeV photons could still plausibly be produced by non-hadronic processes, such as inverse-Compton scattering from a high energy electron population in or near Galactic sources (see e.g., \citep{1989AA...213L..23S, 2011arXiv1103.4284N, 2011A&A...527L...4B, 2012MNRAS.424.2249K}), although there are flux upper limits in the Northern \citep{Chantell1997, Borione1998, Feng_KG_2015, Kang_KG_2015a, Kang_KG_2015b} and Southern \citep{2013PhRvD..87f2002A} hemispheres.
These photon limits, except for \citep{Kang_KG_2015a}, are for energies of order 1 PeV or below, whereas this analysis is most sensitive at energies above 100 PeV.

\acknowledgments

We acknowledge the support from the following agencies:
U.S. National Science Foundation-Office of Polar Programs,
U.S. National Science Foundation-Physics Division,
University of Wisconsin Alumni Research Foundation,
the Grid Laboratory Of Wisconsin (GLOW) grid infrastructure at the University of Wisconsin - Madison, the Open Science Grid (OSG) grid infrastructure;
U.S. Department of Energy, and National Energy Research Scientific Computing Center,
the Louisiana Optical Network Initiative (LONI) grid computing resources;
Natural Sciences and Engineering Research Council of Canada,
WestGrid and Compute/Calcul Canada;
Swedish Research Council,
Swedish Polar Research Secretariat,
Swedish National Infrastructure for Computing (SNIC),
and Knut and Alice Wallenberg Foundation, Sweden;
German Ministry for Education and Research (BMBF),
Deutsche Forschungsgemeinschaft (DFG),
Helmholtz Alliance for Astroparticle Physics (HAP),
Research Department of Plasmas with Complex Interactions (Bochum), Germany;
Fund for Scientific Research (FNRS-FWO),
FWO Odysseus programme,
Flanders Institute to encourage scientific and technological research in industry (IWT),
Belgian Federal Science Policy Office (Belspo);
University of Oxford, United Kingdom;
Marsden Fund, New Zealand;
Australian Research Council;
Japan Society for Promotion of Science (JSPS);
the Swiss National Science Foundation (SNSF), Switzerland;
National Research Foundation of Korea (NRF);
Villum Fonden, Danish National Research Foundation (DNRF), Denmark

\end{document}